\documentclass[twocolumn,longbibliography,10pt,superscriptaddress]{revtex4-2}

\usepackage{amsmath}
\usepackage{graphicx,xcolor}
\usepackage{mathtools}
\usepackage{hyperref}
\usepackage{amssymb}

\usepackage{soul}

\begin{document}

\title{From Stars to Molecules: AI Guided Device-Agnostic Super-Resolution Imaging}

\author{Dominik Va\v{s}inka}
\thanks{Corresponding authors. \newline Dominik Va\v{s}inka: vasinka@optics.upol.cz \newline Miroslav Je\v{z}ek: jezek@optics.upol.cz}
\affiliation{Department of Optics, Faculty of Science, Palack\'{y} University, 17. listopadu 12, 77900 Olomouc, Czechia}

\author{Filip Jur\'{a}\v{n}}
\affiliation{Department of Optics, Faculty of Science, Palack\'{y} University, 17. listopadu 12, 77900 Olomouc, Czechia}

\author{Jarom\'{i}r B\v{e}hal}
\affiliation{Department of Optics, Faculty of Science, Palack\'{y} University, 17. listopadu 12, 77900 Olomouc, Czechia}

\author{Miroslav Je\v{z}ek}
\thanks{Corresponding authors. \newline Dominik Va\v{s}inka: vasinka@optics.upol.cz \newline Miroslav Je\v{z}ek: jezek@optics.upol.cz}
\affiliation{Department of Optics, Faculty of Science, Palack\'{y} University, 17. listopadu 12, 77900 Olomouc, Czechia}



\begin{abstract}
Super-resolution imaging has revolutionized the study of systems ranging from molecular structures to distant galaxies. However, existing super-resolution methods require extensive calibration and retraining for each imaging setup, limiting their practical deployment. We introduce a device-agnostic deep-learning framework for super-resolution imaging of point-like emitters that eliminates the need for calibration data or explicit knowledge of optical system parameters. Our device-agnostic modeling utilizes diverse, numerically simulated dataset encompassing a broad range of imaging conditions, enabling generalization across different optical setups. Once trained, the model reconstructs super-resolved images directly from a single resolution-limited camera frame with superior accuracy and computational efficiency compared to state-of-the-art methods. We experimentally validate our approach using a custom microscopy setup with controllable ground-truth emitter positions. We also demonstrate its versatility on stellar astronomy and single-molecule localization microscopy datasets of point-like sources, achieving high resolution without prior information. Our findings establish a pathway toward universal, calibration-free super-resolution imaging, expanding its applicability across scientific disciplines.
\end{abstract}

\maketitle



\section*{Introduction}

Optical resolution is fundamentally constrained by diffraction. This limits the ability to observe structures comparable to the wavelength of light divided by the numerical aperture of the imaging system, i.e., the famous Rayleigh resolution limit~\cite{Rayleigh1879}. Super-resolution imaging has emerged as a transformative technique in disciplines ranging from biomedical~\cite{Sahl2017, Schermelleh2019} and materials science~\cite{Pujals2019} to astronomy~\cite{Starck2002, Bertero2009}, by circumventing this limit and revealing previously inaccessible structural details. Numerous applications rely specifically on imaging point-like or single-emitter sources, respective to the resolution limit. In biology, single-molecule localization microscopy enables nanoscale visualization of cellular structures~\cite{Rust2006, Betzig2006, Balzarotti2017, Lelek2021, Fiolka2022, Testa2024}. Quantum physics leverages super-resolution for characterizing quantum dots~\cite{Lidke2005} and precisely imaging cold atoms in optical lattices~\cite{McDonald2019, Impertro2023}. Astronomy benefits through resolving individual stars and galaxies~\cite{Cava2017, Zhang2024}.

Despite these successes, super-resolution methods require precise calibration and extensive knowledge of optical parameters. This requirement presents a substantial bottleneck, often involving laborious measurements, calibration data acquisition, and computationally expensive model retraining~\cite{Lelek2021, Aristov2018}. Furthermore, practical applications frequently struggle with inhomogeneities and instabilities in imaging conditions, which further limit the versatility of existing algorithms~\cite{Beckers1993, Guo2024, Qiu2025}.

To address these fundamental challenges, we propose a strategy for the data-driven deep-learning frameworks that is inherently device-agnostic, eliminating the need for calibration or prior system-specific information. Our method exploits numerically simulated data encompassing a wide range of optical conditions, enabling extreme generalization and adaptability across different imaging setups. Once trained, the device-agnostic model delivers rapid and accurate super-resolution reconstructions directly from a single, diffraction-limited image.

In this work, we demonstrate the efficacy of our approach through computational evaluations and experimental validations across multiple benchmarks, outperforming statistical Bayesian approaches and deep-learning-based state-of-the-art methods. We further confirm the practical advantages and broad applicability of our approach across diverse datasets, ranging from localization microscopy to stellar astronomical imaging of point-like sources. The developed framework paves the way towards universal, calibration-free super-resolution imaging, significantly enhancing its accessibility and impact across scientific and technological disciplines. Moreover, the device-agnostic training paradigm is not tied only to imaging and can be applied to any application, where broader generalization ability is a beneficial or even crucial feature.



Super-resolution imaging can be achieved through various approaches, such as linear inverse and statistical Bayesian algorithms~\cite{Bertero2021}, sparse representation~\cite{Yang2010}, tomographic image synthesis~\cite{Luo2015, Fiolka2022, Bianco2023}, and methods based on blinking emitters~\cite{Rust2006, Betzig2006, Balzarotti2017}. Many deep learning super-resolution methods have been developed in the last decade~\cite{Rivenson2017, Nehme2018, Moen2019, Speiser2021, Henriques2021}.
Recently, artificial intelligence has been used to identify novel super-resolution microscopy setups~\cite{Krenn2024}.

Regardless of the specific approach, super-resolution imaging can generally be classified into two categories: reconstruction and parameter estimation. Reconstruction aims to directly restore the whole super-resolved image~\cite{Nehme2018, Bertero2021}, while maximizing the fidelity to the underlying structure. In contrast, estimation focuses on extracting key parameters and features, such as emitter localization~\cite{Smith2010, Speiser2021, Hekrdla2025}. The localization performance, often emphasizing minimal false-positive detections at the cost of information loss, rapidly decreases for images with a higher number of overlapping emitters, which typically limits its application to stochastically blinking or photoactivated samples~\cite{Ovesn2014, TAKESHIMA2018, Holden2011}. In the following discussion, we will focus exclusively on the former: image reconstruction.

Achieving accurate reconstruction of an object based on a single intensity image requires substantial knowledge about the imaging system under consideration. Central to this knowledge is an accurate estimation of the point spread function, which characterizes the response of the system to a point source of light. The reconstruction quality is inherently linked to the accuracy of the point spread function estimate; any inaccuracies will lead to poor reconstruction results~\cite{Booth2015}. However, its precise estimation entails additional calibration measurements involving an isolated emitting point source. This condition can prove demanding, especially in systems with high concentration of emitters, low signal-to-noise ratio, or temporally developing systems. While procedures for in-situ point spread function estimation for room-temperature microscopes exist~\cite{Xu2020, Liu2024}, these methods require an additional calibration step and are not feasible across domains, such as imaging of astronomy objects, cryostat-positioned semiconductor quantum dots, or electron microscopy. Additionally, imaging systems commonly exhibit variations in the shape of the point spread function across the image. Consequently, reconstructing larger areas may require separate calibration for each segment~\cite{Xu2020}.

Some approaches can be adapted to perform a blind reconstruction, operating without explicit prior knowledge of the point spread function~\cite{Campisi2007}. Instead, these methods impose additional assumptions on the imaging system, typically in the form of a predefined point spread function shape. During reconstruction, parameters of the point spread function, such as its width or other defining characteristics, are iteratively estimated directly from the input image. As a result, these methods introduce imperfections by simultaneously reconstructing the observed object and the point spread function of the imaging system. Altogether, blind reconstruction remains a non-convex optimization problem, posing a challenging task in real-life scenarios and often yielding inferior results compared to non-blind algorithms.

\begin{figure}
	\includegraphics[width=\columnwidth]{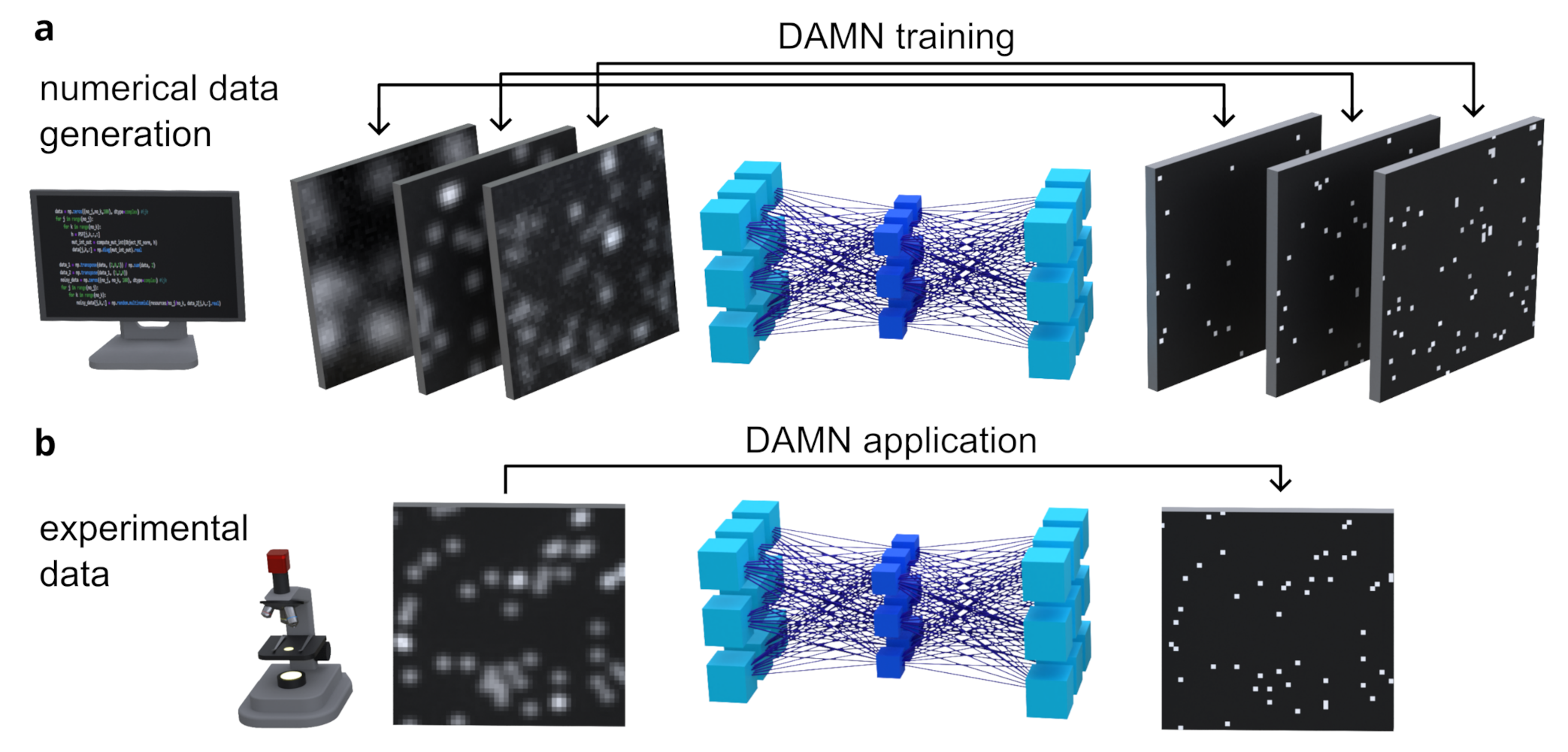}
	\caption{
		Schematic representation of the device-agnostic modeling approach and its application to intensity images of point-like emitting sources. \textbf{a} The model is trained using numerically simulated data pairs comprising resolution-limited noisy images alongside their super-resolved counterparts. Each training sample represents a unique combination of underlying optical parameters, such as the width of the point spread function and the signal-to-noise ratio. \textbf{b} Following training, this model is applied to enhance the resolution of experimental images acquired using a real-life imaging system.
	}
	\label{fig:1}
\end{figure}

Similar device-dependent constraints arise in data-driven deep learning reconstruction. These approaches often outperform classical algorithms in both reconstruction and parameter estimation tasks~\cite{Rivenson2017, Nehme2018, Speiser2021, Henriques2021} and highly improve the processing of large astronomical surveys, as emphasized by the fast inference speed after training~\cite{Schawinski2017, Sureau2020, Lauritsen2021, Dabbech2022, Connor2022, HuertasCompany2023, Akhaury2024, Akhaury2025}. Yet, their calibration poses an even greater challenge. Unlike traditional methods, a data-driven deep learning model infers the reconstruction mapping by extracting relevant information from observed samples. However, due to the device-dependent nature of a typical training process, the model can only be applied to a particular point spread function profile and additional conditions, such as emitter power, background noise distribution, and concentration of emitters, each requiring prior estimation from the calibration datasets~\cite{Belthangady2019}. Modifying these parameters requires remeasuring the calibration set and retraining the model, making it a resource-expensive procedure. Several approaches have moved beyond purely data-driven training by incorporating physical models into the deep learning framework. Examples include plug-and-play methods, which combine learned denoisers with classical iterative reconstruction schemes, and deep unrolling, which embeds the imaging algorithm into a trainable, iteration-inspired architecture~\cite{Li2022DNN, Kamilov2023}. By integrating explicit physical priors, these hybrid approaches can improve robustness and interpretability compared to fully data-driven models. At the same time, the priors make these strategies even more model-dependent and less generalizable. Finally, reconstruction in astronomical observations may benefit from using multiple input frames of the same region~\cite{Sukurdeep2025}, enabling gradual improvement in reconstruction quality with each additional frame~\cite{Hirsch2011}, or from capturing images at different wavelengths~\cite{Koopmans_arxiv}. However, such techniques are beyond the scope of the targeted single-frame domain.



\section*{Results}
\subsection*{Device-agnostic modeling}

We developed a strategy for training device-agnostic neural networks. The framework presented in this study is not tied to a single specific model but can be readily applied to any neural network architecture. We use it to train a device-agnostic modeling network (DAMN) designed to reconstruct intensity images of point-like emitting sources, as illustrated in Fig.~\ref{fig:1}. Unlike traditional approaches, a DAMN model operates using a single intensity frame without requiring additional information about the sample and optical parameters of the imaging system. Without the need for calibration measurements or retraining, it can directly process images from diverse applications. Our model leverages the fully convolutional architecture of a neural network (refer to the Methods section for details), enabling it to reconstruct frames of varying shapes and sizes. Moreover, the reconstruction process is performed by a single-pass propagation through the neural network, resulting in rapid reconstruction of super-resolved images.

We train the DAMN model using pairs of targeted ideal objects and their corresponding resolution-limited images generated through numerical simulation. The ensemble of data pairs covers various optical parameters, such as emitter power and concentration, background noise distribution, and even different shapes and widths of the point spread function, including various asymmetries. Their ranges were selected to cover the majority of realistic situations (additional details in the Methods section). The extent of the synthetic training data exceeds the conventional approaches in the astronomy or single-molecule microscopy domains. Subsequently, we utilized incremental learning techniques~\cite{vandeVen2022} for the model to acquire knowledge about the widest possible parameter combinations. As a result of the optimization and training process, the trained model accurately reconstructs super-resolved images independently of the underlying imaging parameter values. Consequently, without the need for retraining or calibration data, the model can be applied to diverse imaging systems and is robust to their parameter changes, including inhomogeneity across a field of view and instability, such as temporal variations and fluctuations. Moreover, the DAMN model demonstrates a high robustness to optical aberrations as characterized in the Supplementary Aberration analysis section.

\subsection*{Evaluation using simulated datasets}

\begin{figure}
	\includegraphics[width=\columnwidth]{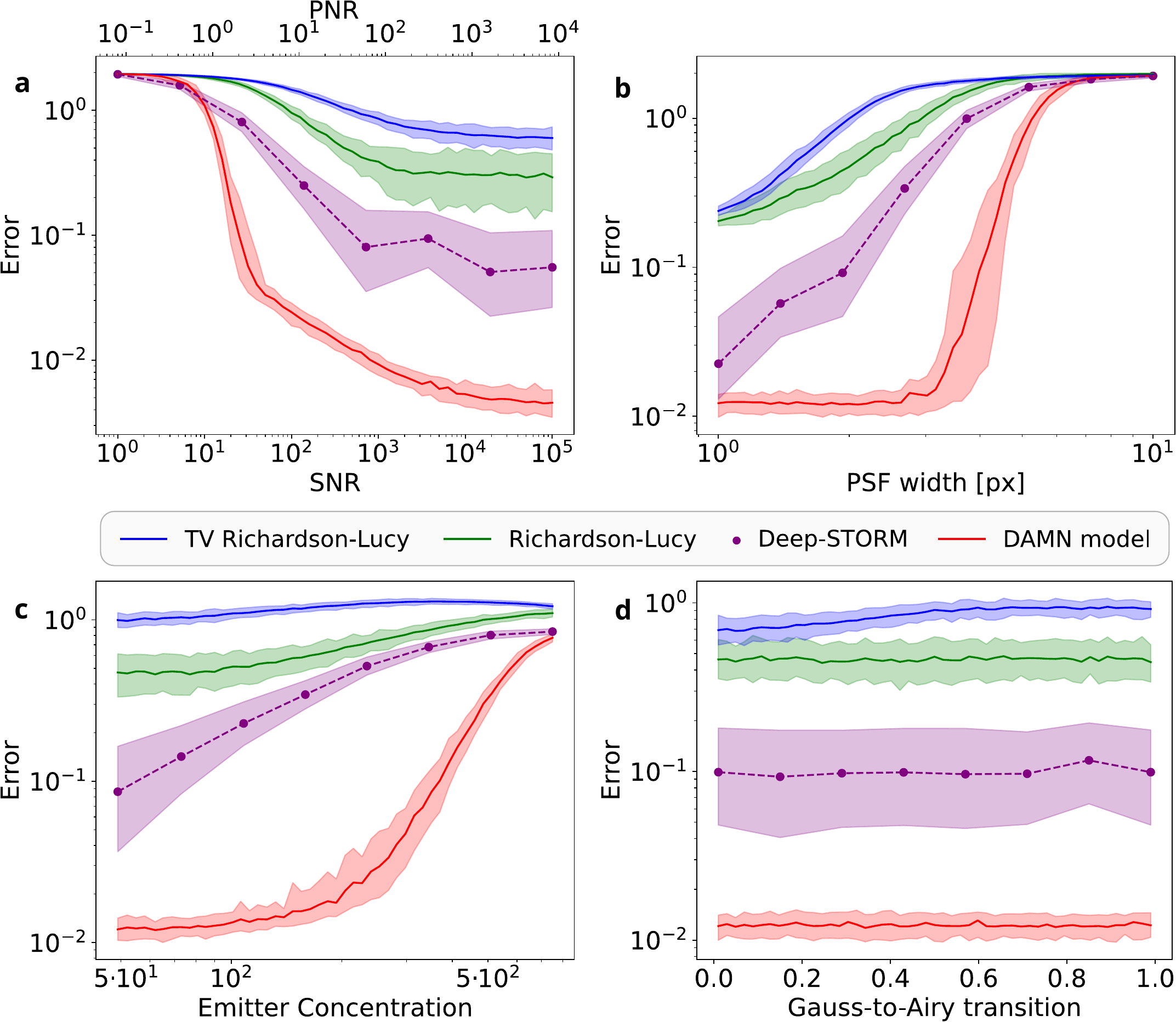}
	\caption{
		Performance of super-resolving methods on simulated data. The dependence of the mean absolute error on \textbf{a} the signal-to-noise ratio (SNR), \textbf{b} the width of a Gaussian point spread function (PSF), \textbf{c} the number of emitters in the image (concentration), and \textbf{d} the continuous transition between a Gaussian and an Airy PSF, respectively. The resulting averages of the Richardson-Lucy algorithm (green), its variant with total variation regularization (blue), individual Deep-STORM neural networks (purple), and the DAMN model (red) are accompanied by their 90\% confidence intervals over the test set. Panel \textbf{a} is provided with a secondary horizontal axis recalculating the SNR values to the peak-to-noise ratio (PNR). Across all panels, the DAMN model consistently outperforms the alternative approaches by up to two orders of magnitude. The optical parameters not investigated in a given panel have the following values: SNR~$= 500$, the average noise intensity~$= 10$, the concentration~$= 50$, and the Gaussian PSF with $\sigma = 2$~px.
	}
	\label{fig:2}
\end{figure}

To characterize the effectiveness of device-agnostic modeling, we conduct a comparative analysis with established reconstruction algorithms. Namely, we implemented the Richardson-Lucy deconvolution based on Bayesian estimation with a uniform prior~\cite{Richardson1972, Lucy1974}, and its variant utilizing a total variation regularization~\cite{Dey2006} (see Methods for details). These widely used iterative algorithms reconstruct images by repeatedly utilizing a known point spread function. We also compare our results against Deep-STORM~\cite{Nehme2018}, a deep-learning method for image reconstruction in localization microscopy (see Supplementary Deep-STORM section). This device-specific convolutional network is trained on simulated data with imaging parameters that precisely match the testing regimes, including point-spread-function profile, signal-to-noise ratio, and emitter density. Therefore, all benchmarking approaches have the advantage of utilizing prior information about the imaging setup.

We evaluate the results over a wide range of optical parameters to demonstrate the device-agnostic generalization ability using separate test sets of image pairs not included in the training data. We quantify the performance of all methods using the mean absolute error between the reconstructed image and its target object, averaged across each test dataset. In contrast to both Richardson-Lucy algorithms and Deep-STORM, which require (and were supplied with) prior knowledge of the imaging system, our DAMN model operates independently of any device-specific information. Despite this distinction, the device-agnostic model (without prior information) outperforms the alternative approaches in terms of reconstruction quality, as discussed in the following text.

Panels a-c in Fig.~\ref{fig:2} illustrate the dependence of mean absolute error on the emitter power, the width of the point spread function, and the concentration of emitters, respectively. The results of these log-log graphs are characterized using an average value over the test dataset with a corresponding 90\% confidence interval. Panel a contains dual horizontal axes representing the varying emitter power using both the signal-to-noise ratio (SNR) and the peak-to-noise ratio (PNR). For additional details regarding the explored parameters and evaluation of both approaches, see the Methods section. The DAMN model, represented by the red color, exhibits significantly better performance than the blue and green-colored Richardson-Lucy algorithms and even surpasses the purple Deep-STORM. Similar results can also be observed in panels b and c. Panel b depicts the error dependence on the point spread function width (consult the Methods section). The remaining panel c characterizes the dependence on the emitter concentration, referring to the collective number of emitters in a single image. As seen across all three panels, our device-agnostic approach provides more accurate reconstructions than the alternatives across all testing regimes. Moreover, each Deep-STORM graph point corresponds to an individually trained neural network that utilizes precise knowledge of the test imaging parameters. The depicted errors were evaluated across the whole image and can be, alternatively, expressed as per-pixel errors by dividing the values by the number of pixels.

The panel d in Fig.~\ref{fig:2} visualizes the error dependence on the point spread function shape, gradually transitioning from a Gaussian profile to an Airy pattern of the same width. As seen from this semi-logarithmic graph, the performance of all methods remains approximately constant during this transition. Such behavior is expected from the device-specific approaches, as we always provide them the correct point spread function profile. On the other hand, the DAMN model was trained solely on the exact Gaussian and Airy point spread functions, each constituting approximately half of the dataset. Despite this simplification in the training process, the graph unambiguously demonstrates the adaptive ability of the DAMN approach to generalize to previously unseen point spread function shapes. Moreover, the DAMN model exhibits strong robustness against optical aberrations in the image (as discussed in the Supplementary Aberration analysis section), despite never seeing them during training. This adaptability is beneficial, as it significantly reduces the data and simulation complexity required for training device-agnostic models. Altogether, the presented panels demonstrate numerous benefits of the DAMN approach, proving it superior to the established algorithms.

\subsection*{Optical microscopy experimental validation \ \ \ \ \ \ with partial control over ground truth}

\begin{figure}
	\includegraphics[width=0.8\columnwidth]{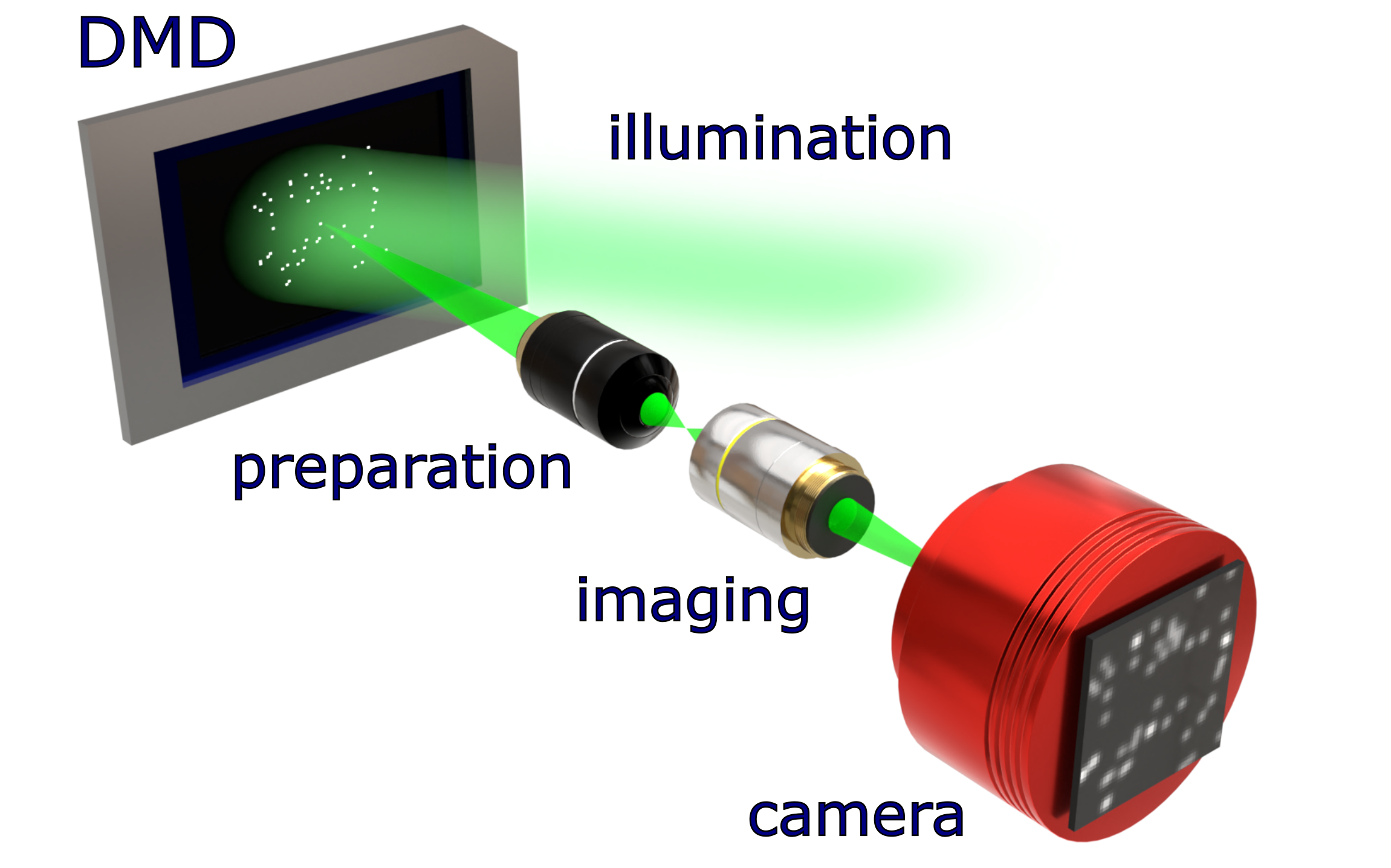}
	\caption{
		Schematic illustration of the optical setup used to collect experimental data pairs. A ground-truth mask is imposed on the digital micromirror device (DMD) by configuring its mirrors. An incoherent illumination light reflected by these mirrors impinges a high-resolution preparation system. The DMD-imposed mask is re-imaged into the front sample plane of the preparation system, creating point-like emitters with the intended spatial distribution. The imaging part of the setup, comprised of a low-resolution microscope objective, images the sample-plane emitters onto a camera. The resulting camera-captured intensity image and the DMD-imposed mask represent the experimental data pairs.
	}
	\label{fig:3}
\end{figure}

To validate our approach beyond simulated data, we applied it to reconstruct experimentally acquired images. Such data inherently lack ground truth information, which prevents the use of most quantitative evaluation metrics. To bridge the gap between purely simulated and real-world, but fully uncontrolled, experimental data, we developed a custom-built optical microscopy setup, illustrated in Fig.~\ref{fig:3} and shown in detail in Supplementary Fig.~1. Unlike the traditional super-resolution benchmarking, this microscope provides partial control over the spatial distribution of emitting point-like sources and complete ground-truth information about their positions. An incoherent light illuminates a digital micromirror device (DMD), onto which a mask representing the targeted ground-truth is imposed. By configuring the micromirrors, we prepare a mask containing the targeted number of sources positioned at desired locations. As such, the control is limited to activating emitting sources at discrete grid positions. This mask is then re-imaged by a high-demagnification and high-resolution preparation system, creating point-like emitters and thus forming a sample object. Subsequently, a low-resolution imaging microscope projects the emitters onto a camera. The resulting intensity profile captured by the camera serves as the resolution-limited input for reconstruction methods. Together with the ground-truth mask, this setup provides experimental test data pairs, which allow for exact metric quantification of each reconstruction method performance. The collected dataset does not fully replicate or substitute real microscopy conditions due to its restriction to two-dimensional grid distributions and uniform illumination. However, it represents an essential link and intermediate middle-point between idealized simulations with complete control and knowledge over ground truth and fully experimental data lacking both ground truth knowledge and control. For further details on the optical setup, see the Supplementary Experiment section.

\begin{figure}
	\includegraphics[width=\columnwidth]{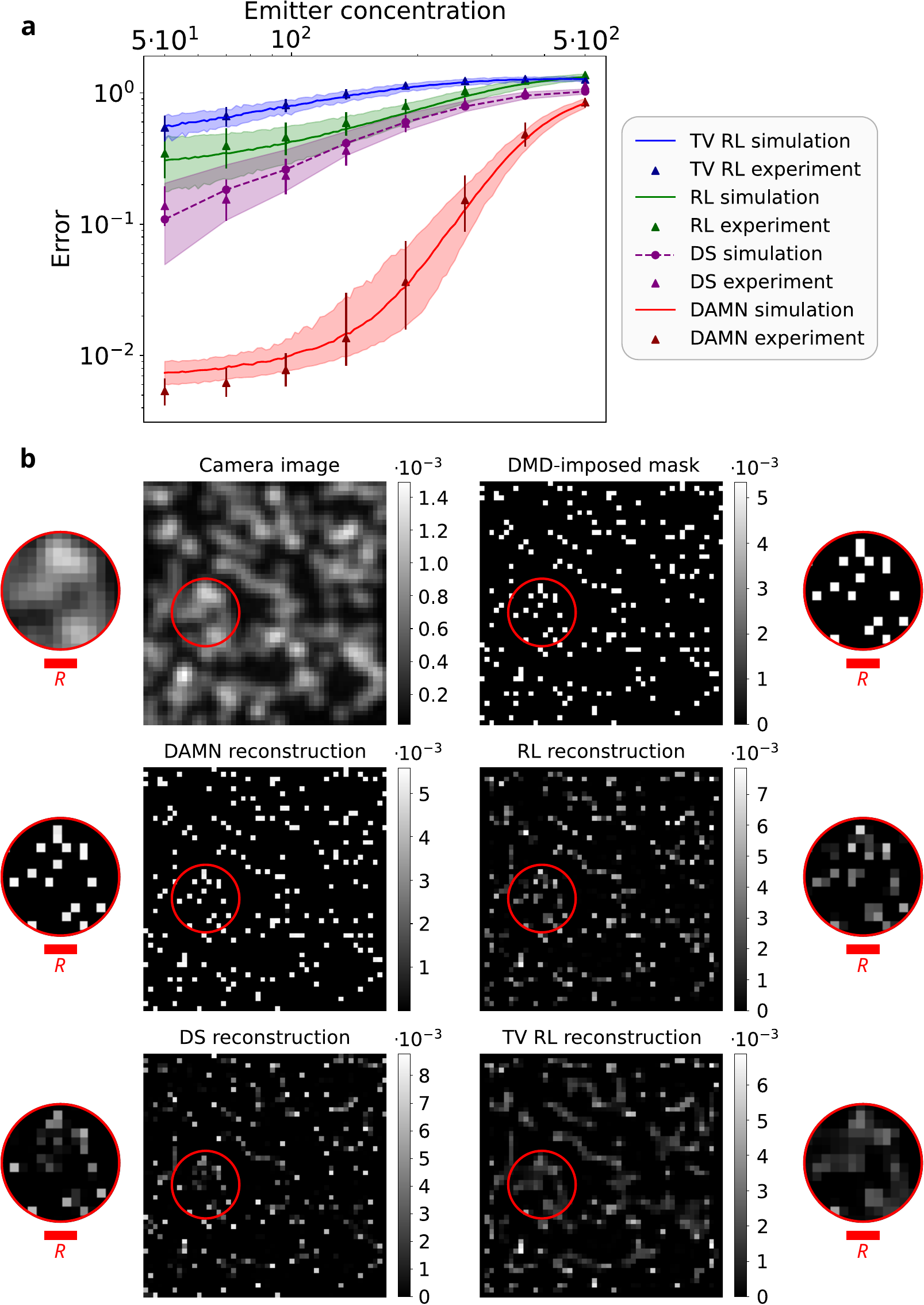}
	\caption{
		Performance of super-resolution methods on experimental data with the partial control and complete knowledge of ground truth. \textbf{a} The colored dots represent the mean absolute error between the DMD-imposed masks and the super-resolved images reconstructed by each method: Richardson-Lucy (RL, blue), Total variation Richardson-Lucy (TV RL, green), the device-agnostic model (DAMN, red), and individual Deep-STORM models (DS, purple). These errors were evaluated across various emitter concentrations with their 90\% confidence intervals. The accompanying continuous lines depict error values derived from simulated data using optical parameters estimated for our imaging system. \textbf{b} A typical camera image containing nearly 200 emitters, alongside its corresponding DMD-imposed mask and reconstructions from each method. The circled areas contain a magnified region for easier visual comparison. It is evident that the DAMN model significantly outperforms the alternative approaches, even in regions where the mutual emitter distance is well below the Rayleigh resolution limit $R = 3.9$~px (inset scale bars).
	}
	\label{fig:4}
\end{figure}

Using this setup, we investigated the performance of all methods while varying the concentration of emitters. Panel a in Fig.~\ref{fig:4} illustrates the mean absolute error between reconstructed images and their corresponding ground-truth masks using the experimental data, shown as dot markers. For comparison, the continuous lines represent the results obtained using simulated data with corresponding optical parameters. For Deep-STORM, results on simulated data are also illustrated as scatter points, since each requires a separately trained neural network. The DAMN model outperforms both Richardson-Lucy deconvolutions by orders of magnitude. Similarly, it achieves better reconstruction performance than the device-specific Deep-STORM, even when applied to images acquired from experimental measurements. This improvement is achieved despite operating significantly below the Rayleigh resolution limit of our imaging system. To further emphasize these findings, panel b in Fig.~\ref{fig:4} shows a representative experimental image containing nearly 200 emitters and its reconstruction by each method. As observed, the DAMN model provides a near-perfect reconstruction of the original image without any calibration. The two closest resolved emitters are 1 px apart, even in dense clusters, which is a distance almost 4 times below the Rayleigh resolution limit, $R = 3.9$~px. (The resolution improvements for reconstructions with an upsampling DAMN model are evaluated in the following section.) In contrast, the other methods exhibit visible deviations and artifacts even with full access to the imaging parameters. This experimental demonstration highlights the benefits of our device-agnostic modeling approach, which significantly outperforms the established algorithms that rely on prior knowledge of imaging parameters.

\subsection*{Stellar astronomy and localization microscopy demonstration}

Since real applications benefit from the highest possible reconstruction precision, we also applied the device-agnostic modeling to train an additional network with upsampling layers in its architecture (see Methods). This DAMN model inherently provides an eightfold increase to reconstructed spatial dimensions, allowing reconstruction on a finer image grid. In this section, we used this upsampling model on three distinct experimentally acquired datasets of point-like sources ranging from stellar astronomy to single-molecule microscopy. All presented results were obtained without any recalibration or retraining of the DAMN model. Moreover, no pre-processing was applied to the measured images.

\begin{figure}
	\includegraphics[width=0.9\columnwidth]{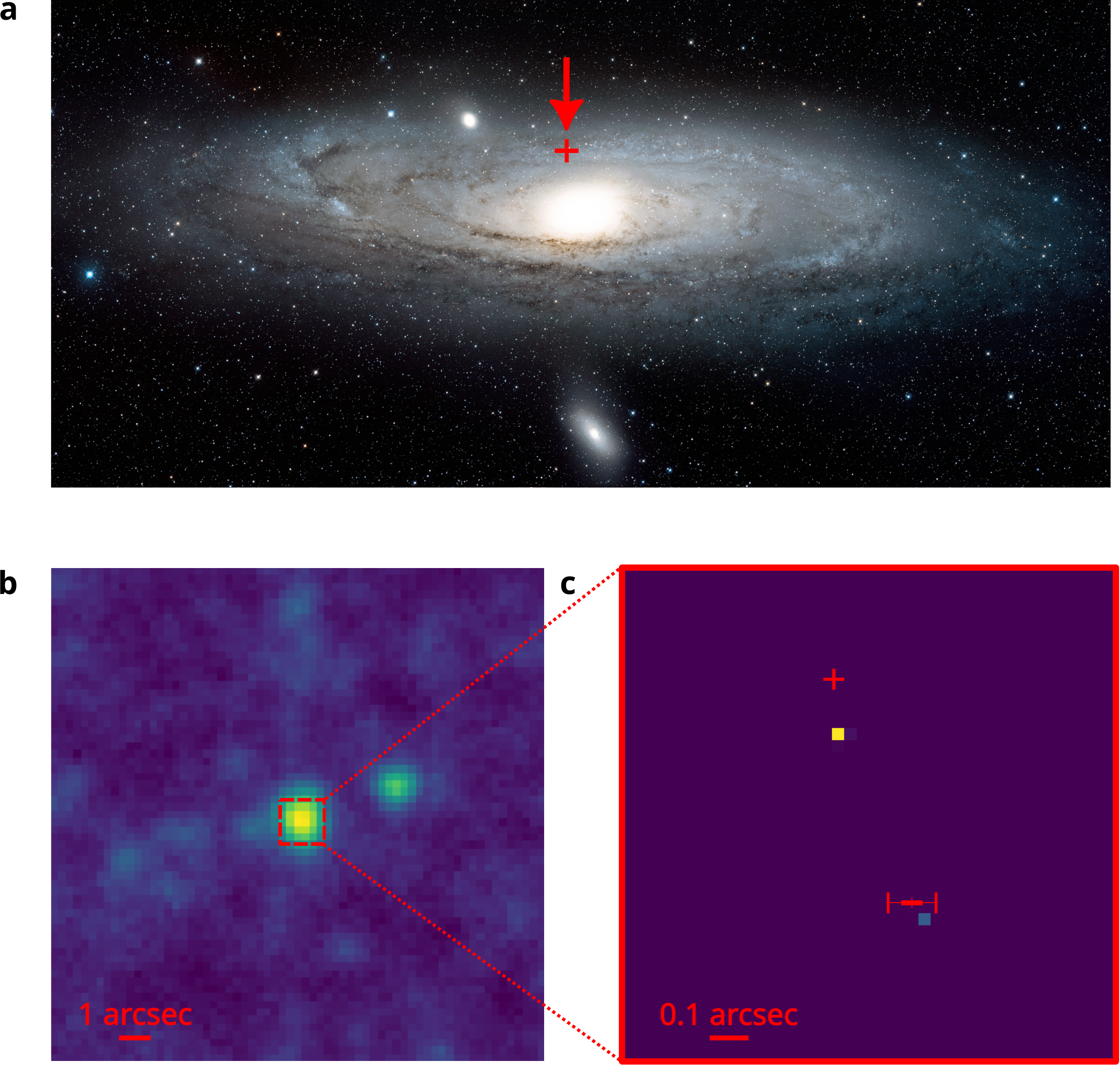}
	\caption{
		Demonstration of the super-resolving capabilities of the DAMN model on an astronomical image of a dense stellar field. \textbf{a} Contextual wide-field view of the Andromeda galaxy (M31) by ESA/Hubble~\cite{M31_galaxy}, rotated and cropped. The location of the analyzed region is indicated. \textbf{b} Zoomed-in outer field of M31 obtained with the ground-based Pan-STARRS1 telescope, showing an unresolved double-star system in the center. \textbf{c} High-resolution reconstruction of the same region produced by the DAMN model, further zoomed for visualization. Red markers indicate the star positions and their standard deviation uncertainties derived from Gaia DR3 astrometric data, based on the Gaia space-based observatory. Notably, one uncertainty is significantly larger as that star magnitude borderlines with Gaia sensitivity. The reconstruction does not use any calibration data or prior information on the employed imaging system.
	}
	\label{fig:5}
\end{figure}

First, we applied the DAMN model to a ground-based image of the outer field of the Andromeda galaxy (M31). The image was obtained with the Pan-STARRS1 wide-field survey telescope~\cite{Chambers2016PS1}. Within the field of view, a double-star system was identified using astrometric data from the Gaia Data Release 3~\cite{GaiaCollaboration2023DR3}, based on the Gaia space observatory, which provides sub-pixel positional information for both components. The angular separation of the two stars, approximately 0.6~arcsec, lies well below the approximate 1.4~arcsec Rayleigh resolution limit of the Pan-STARRS1 imaging system. Consequently, the double-star system is fully unresolved in the original ground-based image, as shown in Fig.~\ref{fig:5}~b. Nonetheless, the DAMN model, having no explicit knowledge of the imaging parameters, still resolved the two distinct point-like sources, despite operating at nearly 2.5~times below the resolution limit. Panel~c depicts the reconstructed region, zoomed to enhance visualization of the double-star system. As shown, both stars are clearly resolved, with an estimated angular separation of 0.52~arcsec. Therefore, the error of the double star distance reconstruction is approximately 0.08~arcsec, which is 17.5~times below the Rayleigh resolution limit. Moreover, the ratio of reconstructed intensities is 3.46, which is comparable to the targeted 3.65 based on Gaia. This result demonstrates the ability of the DAMN model to accurately and precisely recover sub-diffraction structures from experimentally acquired images. Fig.~\ref{fig:5}~a additionally shows a contextual wide-field image of the Andromeda galaxy by ESA/Hubble~\cite{M31_galaxy} with the location of the double-star system indicated.

\begin{figure*}
	\includegraphics[width=0.98\textwidth]{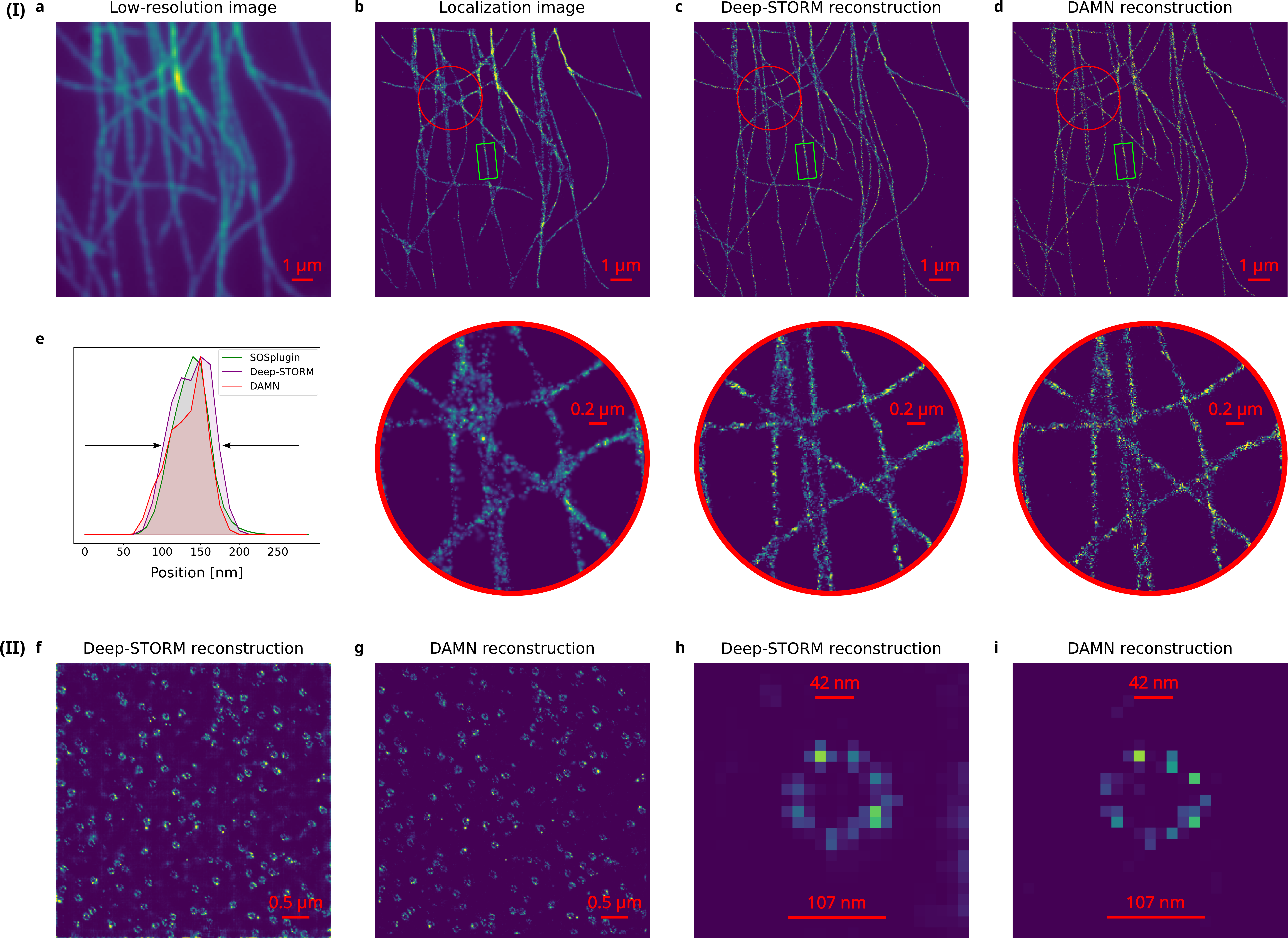}
	\caption{
		Demonstration of the DAMN model on \textbf{(I)} a high-concentration tubulin dataset from the single-molecule localization microscopy challenge, and \textbf{(II)} a nuclear pore complexes dataset.
		\textbf{(I) a} A low-resolution image integrated from the set of 500~measured images, containing numerous emitting molecules. \textbf{b} A reconstruction rendered from a localization table provided by the SOS Plugin. \textbf{c} A reconstruction provided by Deep-STORM, calibrated using detailed information on the imaging setup. \textbf{d} A super-resolved image produced from the calibration-free, device-agnostic model without prior information. Panels \textbf{b-d} include inset images with a magnified region highlighting the resolution details. \textbf{e} Projection of the microtubule profile over the green rectangle segment. The attained full width at half maximum was 56~nm for SOS Plugin, 73~nm for Deep-STORM, and 55~nm for the DAMN model.
		\textbf{II} Reconstruction of the nuclear pore complex dataset containing Nup96 labeled with BG-AlexaFluor647. \textbf{f} Large field-of-view reconstruction generated from the 50,000 frames produced by Deep-STORM, showing multiple nuclear pore complexes. \textbf{g} Corresponding large field-of-view reconstruction obtained with the DAMN model. \textbf{h} Close-up of a single nuclear pore complex reconstructed by Deep-STORM from a subset of 5,000 drift-free frames. \textbf{i} Close-up of the same nuclear pore complex reconstructed by DAMN. Panels \textbf{h} and \textbf{i} also indicate the expected ring diameter and nearest-neighbor spacing of the characteristic octagonal geometry for reference.
	}
	\label{fig:6}
\end{figure*}

We next applied the same DAMN model, without retraining, to a publicly available dataset of tubulin structures~\cite{SMLMRepository, Sage2015}, which was experimentally acquired for the single-molecule localization microscopy challenge. As such, the ground-truth emitter positions and intensities are not available for direct quantitative evaluation. Therefore, we assessed reconstruction quality by comparing it with established methods. Fig.~\ref{fig:6}~(I) presents the performance evaluation using these 500 tubulin images ($128 \times 128$ pixels) with high emitter density. Panel a shows the corresponding integrated low-resolution image. Panel b shows a reference reconstruction provided by the SOS Plugin~\cite{Reuter2014}, a least-square localization method assuming a Gaussian point spread function, which participated in the challenge and is available as a plugin for ImageJ. In this reconstruction, each localized emitter is rendered as an upsampled Gaussian profile whose width corresponds to the localization uncertainty. Panel c depicts a tubulin reconstruction obtained with Deep-STORM, using a model specifically fine-tuned and trained for this dataset by the original authors~\cite{Nehme2018}. Both reference approaches are device-specific and require additional prior information about the imaging parameters. In comparison, the DAMN model directly processes the dataset without calibration or prior knowledge and produces the super-resolved tubulin image, see panel d. The highlighted areas (red) show a close-up comparison of all three approaches.

Both DAMN and Deep-STORM accurately recover the underlying tubulin structure even in densely populated regions, achieving visual quality significantly surpassing that of the SOS Plugin localization method. For example, two vertically oriented, overlapping microtubules in the insets are clearly resolved by both deep-learning techniques. Moreover, a close inspection reveals that the DAMN reconstruction exhibits a sharper, more confined microtubule structure than Deep-STORM. Panel e visualizes the projection of a single microtubule profile over the green segment. For each method, we evaluated the full width at half maximum of the reconstructed microtubule profile as 56~nm for the SOS Plugin, 73~nm for Deep-STORM, and 55~nm for the DAMN model. These results indicate that the DAMN approach combines robust performance in high-density conditions with a spatial precision comparable to localization-based methods. Overall, this comparison highlights the reconstruction capability of device-agnostic models for challenging microscopy applications.

Finally, since no ground-truth information is available for the tubulin dataset, we also evaluated the methods using microscopy data with partial structural ground truth: nuclear pore complexes, whose architecture is well-characterized. Specifically, we utilized a dataset of the scaffold nucleoporin Nup96 labeled with BG-AlexaFluor647~\cite{Thevathasan2019}. The nuclear pore complexes exhibit eightfold rotational symmetry, with Nup96 expected to occupy positions on a ring of approximately 107~nm in diameter and a nearest-neighbor spacing of 42~nm, given the labeling geometry. These distances are approximately 3.4~times and 8.6~times smaller than the Rayleigh resolution limit of 360~nm of the employed imaging setup, respectively. Therefore, the dataset represents a challenge for super-resolution reconstruction. Using 50,000 low-resolution frames, Fig.~\ref{fig:6}~(II) compares reconstructions obtained with Deep-STORM, trained using the estimated imaging parameters, and with the DAMN model. While both approaches successfully recover the octagonal arrangements, Deep-STORM exhibits stronger background noise across the field of view.

Moreover, a closer inspection of the dataset reveals a noticeable lateral drift over time, which could, in principle, be compensated for. However, since we want to compare the reconstruction performance of each method without additional post-processing, we instead selected a temporal subset of frames with negligible drift. Accordingly, we include panels showing a single nuclear pore complex reconstructed from 5,000 consecutive frames. While the DAMN model provides a sharper image than the Deep-STORM, the recovered structures are consistent with the known geometry for both cases. Additionally, for the DAMN reconstruction, the estimated ring diameter is 100~nm, and the inter-emitter spacing is 38~nm, corresponding to distance errors of 7~nm and 4~nm, respectively. These results are the ultimate demonstration of the DAMN super-resolving ability. For comparison, the Deep-STORM reconstructed ring diameter is 98~nm, and the inter-emitter spacing is 38~nm, with the distance errors of 9~nm and 4~nm, respectively. We also evaluated the SOS localization plugin on this dataset. However, because it could not reliably localize a substantial fraction of emitters under these conditions, the resulting reconstruction quality was notably inferior to that of both deep-learning approaches. Therefore, it is not included in the figure.



\section*{Discussion}

We demonstrated a device-agnostic approach to super-resolution imaging of point-like emitting sources that utilizes deep learning techniques. The presented approach solves the long-standing problems of measuring calibration data and estimating parameters accompanying the training of a deep learning model. Using only numerically simulated data, we develop a model capable of reconstructing intensity images using a single frame without explicit knowledge of the imaging system. This model can be applied to images of arbitrary shape and size originating from various optical systems, and its calibration-free nature provides robustness against non-uniformity and temporal variations in the optical parameters, including optical aberrations in the image. We compared this device-agnostic model with well-established reconstruction algorithms, namely the classical and regularized Richardson-Lucy deconvolution algorithm, and the state-of-the-art deep learning Deep-STORM. The analysis results clearly demonstrate the superior performance of our approach, which outperforms benchmark methods by up to two orders of magnitude in mean absolute error. Moreover, we designed an advanced optical setup for acquiring experimental images of emitting sources together with their precise ground-truth references. Providing partial control and complete knowledge over the spatial distribution of emitters, this setup enables exact performance quantification using measured data, which, to the best of our knowledge, is unprecedented in the super-resolution imaging domain. Using these experimental data pairs, we verified the superior performance of our approach.

To further demonstrate the universality of our approach, we applied the model to astronomy and microscopy data of point-like sources, achieving significant improvements in resolution in both domains. We reconstructed high-resolution images of tubulin and nuclear pore complexes from single-molecule localization microscopy data. In astronomy, we enhanced the resolution of an image of a fully unresolved double-star system, accurately resolving both its components. Notably, all DAMN reconstructions were performed without any prior knowledge of the optical setup, data preprocessing, or parameter estimation. Despite this, the obtained results surpass those of the established classical and deep learning methods. With sufficient computation resources, even further improvements can be achieved by incorporating more complex data simulations and larger network architectures. By taking this first step, our work lays a foundation for universal image reconstruction, entirely independent of optical settings, thus contributing to the advancement of image processing in various fields.

The device-agnostic paradigm presented in this study is not tied to a single specific neural network. It can be readily applied to any model and architecture that would be beneficial for a given application. Moreover, although the presented work focuses on super-resolution image reconstruction, the same framework can be readily applied to localization tasks, i.e., directly predicting emitter positions and intensities. Such an approach would be particularly beneficial for those single-molecule localization microscopy applications that require explicit information about emitter positions. Finally, the framework is not inherently restricted to optical imaging. The same principles can be applied to virtually any application that would benefit from improved generalization ability and hardware parameter independence.

During the preparation and review of the presented manuscript, we further developed the device-agnostic modeling framework toward dedicated applications in single-molecule fluorescence microscopy~\cite{Dostalova_arxiv} and material engineering for single-frame imaging of quantum dots~\cite{Vasinka_arxiv}. In the former, we extend the analysis using our own experimentally acquired molecular dataset and provide a more detailed comparison with localization-based approaches. The latter applies the device-agnostic approach to our measured data of high-density In(Ga)As quantum dots and strain-induced dots in 2D monolayer WSe$_2$. Both these studies build directly upon the general framework introduced here and focus on application-specific refinements and analysis.



\section*{Methods}\label{sec:Methods}

\subsection*{Deep Learning Model}

The central contribution of this work is not a specific trained network or a specialized architecture, but the device-agnostic modeling paradigm itself. This framework can be applied to any network with an arbitrary architecture, provided it has sufficient representational capacity and complexity. To emphasize and demonstrate this feature, we trained two distinct architectures, commonly used in deep-learning-based image reconstruction: a fully convolutional neural network, ConvNet (applied to simulated data and optical experiment), that preserves the input spatial dimensions, and a ResNet-based architecture incorporating residual connections and an inherent eightfold upsampling (applied to astronomy, aberrations, and biological samples). Both networks can directly process images of arbitrary size and aspect ratio without retraining.

The fully convolutional model comprises 35 hidden layers with 71 channels each and a final output layer with a single channel. Information flows through each hidden layer via trainable $7 \times 7$ filters that apply local convolutional operations, allowing the network to detect localized patterns. These transformations are followed by a LeakyReLU activation function with a 0.05 negative slope, which introduces a computationally efficient nonlinearity without causing the dying ReLU problem~\cite{Maas2013}. Since we normalize the input samples to unit sum before processing, the output layer utilizes a softmax activation function~\cite{Goodfellow2016} to preserve this condition and ensure non-negative values. To prevent overfitting, each hidden layer is paired with dropout regularization at a 0.01~rate, randomly setting a fraction of input units to zero during training~\cite{Srivastava2014}. This regularization encourages the network to learn robust features that do not rely on any specific neurons. Altogether, this model contains over 8 million trainable parameters and, after initialization, takes approximately 40~ms and 250~ms to process a $100 \times 100$ px and $500 \times 500$ px image, respectively, using an NVIDIA RTX A5000 GPU, without the need for retraining between applications.

The architecture of the upsampling ResNet-based model differs substantially from the dimension-preserving ConvNet. Its basic building block consists of two convolutional layers with 64 channels and a LeakyReLU activation (negative slope of 0.1). Each convolution is followed by batch normalization, which stabilizes the training by normalizing the mean and variance of activations across the batch. In addition, each block incorporates a residual connection, providing an additive shortcut to assist gradient propagation. Three such residual blocks are stacked to form a processing segment operating at a fixed spatial dimension. The full network comprises four of these segments, separated by bilinear interpolation layers that each double the spatial dimensions, resulting in an overall eightfold upsampling. The convolutional filter sizes within the four segments are $5 \times 5$, $7 \times 7$, $9 \times 9$, and $11 \times 11$ pixels, respectively. Owing to the progressive upsampling, the largest $11 \times 11$ filters correspond to an effective receptive field of approximately 1.5 pixels in the original input domain. The output layer produces a single-channel reconstruction and uses a softmax activation to ensure normalized, positive outputs. In total, the upsampling architecture contains more than 7 million trainable parameters and, after initialization, takes approximately 1.8~s and 3.5~s to process a $100 \times 100$ px and $500 \times 500$ px image, respectively, using an NVIDIA RTX A5000 GPU, and does not need to be retrained across applications.

The presented model architecture configurations result from extensive manual optimization across the device-agnostic regime. Each model was trained incrementally on approximately 3 million simulated low-resolution images, with a 25\% validation set split. Incremental learning techniques gradually present the training data to the network. These techniques involve training the model using randomly generated data, which are continuously replaced with newly generated samples throughout the training process~\cite{vandeVen2022}. The level of network complexity and dataset size exceeds the conventional algorithms in the field of emitter reconstruction and localization. Nevertheless, the model remains relatively modest compared to modern large language models, allowing additional potential for improvement, including larger upsampling. Unfortunately, our ability to further increase model complexity is limited by the computational resources available to us.

The training of the DAMN model follows the backpropagation algorithm, which computes the gradient of the loss function with respect to each weight in the network using the chain rule. We use a mean squared error (MSE) loss, calculated between the predicted $I_1$ and target $I_2$ images and averaged over a batch of data samples. Given a set of two-dimensional image pairs, we calculate the loss as
\begin{eqnarray}
	\text{MSE} = \frac{1}{J} \sum_{j=1}^{J} \sum_{k=1}^{K} \sum_{l=1}^{L} 
	\left[ I_1\left( j,k,l \right) - I_2 \left( j,k,l \right) \right]^2,
\end{eqnarray}
where $J$ is the batch size, $K$ and $L$ are the image dimensions, and $I(j,k,l)$ is the intensity of the $j$-th image in the batch at the pixel position $(k,l)$. The gradients obtained from this loss are used to iteratively update the weights with each batch of 128 training samples. To improve convergence towards the minimum loss, we implement the Adam optimizer~\cite{Adam}, which incorporates adaptive moment estimation. Adam updates the weights using both the first and second moments of the gradient, making the descent more efficient. A portion of the training data is reserved as a validation set to actively monitor the convergence using the mean absolute error (MAE) metric
\begin{eqnarray}
	\text{MAE} = \frac{1}{J} \sum_{j=1}^{J} \sum_{k=1}^{K} \sum_{l=1}^{L} 
	| I_1\left( j,k,l \right) - I_2 \left( j,k,l \right) |.
\end{eqnarray}
The metric provides an additional measure of model performance that is, unlike a loss function, not directly optimized during training. Additionally, for the upsampling model operating on the finer grid, the $I_1$ and $I_2$ images are first convolved with a small Gaussian filter to improve convergence during training, and the loss function includes a weak entropy regularization term to promote sparsity.

Any further details about the architecture designs and their training can be found in our GitHub repository~\cite{Github}.
\\


\subsection*{Data Simulation}

A resolution-limited image can be characterized by its underlying optical parameters. In our case, these parameters include the shape and width of the point spread function, the background noise intensity, the emitter power representing the number of emitted photons, and the number of emitters present within a~$50 \times 50$ pixels field of view of an image, which we term emitter concentration. Consequently, all possible resolution-limited images form a high-dimensional space. To implement the device-agnostic framework with a data-driven deep learning algorithm, the model must observe data samples covering the space during training. Therefore, the approach highly benefits from using simulated data pairs, as collecting a sufficient amount of experimentally acquired samples would be exceptionally time-consuming. Additionally, data simulation allows using incremental learning techniques (see Deep Learning Model), which are ideal for applications with large datasets. 

To train the non-upsampling model, the following ranges of optical parameters were used for the data simulation: the average emitter power~$\in \left[ 1, 10^5 \right]$, the average shot noise~$\in \left[ 1, 100 \right]$, the emitter concentration~$\in \left[ 5, 500 \right]$, and the point spread function width~$\in~\left[ 10^{-0.25}, 10^{1.25} \right]$~px~$\approx \left[ 0.5, 17.75 \right]$~px. These ranges represent extreme-to-extreme boundaries to demonstrate the device-agnostic abilities on a difficult, large-scale generalization task.

Training of the upsampling model is significantly more computationally expensive. To alleviate the training process on our limited resources, we chose to slightly narrow the ranges, disregarding the pathological and near-perfect cases. As such, the following ranges were employed: the average emitter power~$\in \left[ 1, 10^4 \right]$, the average noise~$\in \left[ 1, 100 \right]$, the emitter concentration~$\in \left[ 1, 100 \right]$, and the point spread function width~$\in \left[ 10^{0.1}, 10^{1} \right]$~px~$\approx \left[ 1.25, 10.0 \right]$~px. Nonetheless, these ranges still cover a broader range of optical conditions than those typically encountered in practical application.

The generation process of a single simulated data sample follows these steps. First, we generate the emitter concentration, followed by assigning each emitter its power and pixel position in the image. Next, we perform a convolution with the point spread function to simulate the effects of finite resolution. For the data samples utilized by the upsampling model training, we now sum each distinct $8 \times 8$ pixel region. Subsequently, shot noise is added to each pixel of the image. With knowledge of the average shot noise intensity and emitter power, we can calculate the signal-to-noise ratio of the generated sample as $\text{SNR}=\frac{\text{average emitter power}}{\text{average shot noise intensity}}$. Alternatively, we can calculate the peak-to-noise ratio, $\text{PNR}=\frac{\text{peak emitter intensity}}{\text{average shot noise intensity}}$, using the maximum value of the convolved emitter intensity.

To perform the convolution, we generate the convolution matrix of the point spread function using two distinct parameters - shape and width. The shape parameter distinguishes between an Airy pattern $A$, simulating low numerical aperture scenarios, and a two-dimensional Gaussian pattern $G$, typically used for cases with higher numerical aperture values~\cite{Stallinga2010}. The inclusion of both distinct shapes is important for unlocking the high generalization ability. When we originally trained only on one of the shapes, the generalization of the resulting model was inferior compared to that of the model trained on both shapes. Additionally, half of the generated point spread function profiles account for asymmetries by applying an additional squeezing along either the horizontal, vertical, diagonal, or anti-diagonal axis. The convolution matrix is normalized to a unit sum to conserve the emitter power. The pattern functions before squeezing can be expressed as
\begin{eqnarray}
	A ( r, \sigma ) \sim \left[\frac{\text{J}_1(\frac{2r}{\sigma})}{\frac{r}{\sigma}} \right]^2, \\
	G ( r, \sigma ) \sim \text{exp} \left( - \frac{r^2}{\sigma^2} \right),
\end{eqnarray}
where $r$ is the distance from the center, and $\text{J}_1$ is the first-order Bessel function of the first kind. The variable $\sigma$ is the width parameter, which dictates the full width at half maximum of the point spread function as $2 \sigma \sqrt{\ln{2}}$.

It is worth noting that the simulation process can be further enhanced by incorporating additional features and parameters, such as different noise types and optical aberrations~\cite{Li2017, Mik2019}. For this showcase study, we opted to simplify the simulation to demonstrate the strong generalization capability stemming even from the simplified data and to reduce the computational demand during training. Despite this simplification, the DAMN model achieves excellent results across all tested use cases and demonstrates significant generalization ability beyond the expected regime, as shown in Fig.~\ref{fig:2}~\textbf{d}, and by the already high robustness to optical aberrations, as characterized in Supplementary Fig.~2 and its dedicated section.

Any further details about the data simulation process can be found in our GitHub repository.~\cite{Github}


\section*{Data availability}

The data used in this study are publicly available in our GitHub repository~\cite{Github}.

\section*{Code availability}

The code supporting the results of this study are publicly available in our GitHub repository~\cite{Github}.

\section*{Author Contributions}

M.J. conceived the idea of a device-agnostic approach to image super-resolution and supervised the project. D.V. developed deep-learning models and performed numerical simulations and data processing. F.J. and J.B. developed the optical experiment, performed data acquisition, and participated in data processing. J.B. supervised the experimental part of the project. D.V. wrote the manuscript, and all authors were involved in revising the manuscript.

\section*{Acknowledgments}

We acknowledge the massive use of cluster computing resources provided by the Department of Optics, Palack\'{y} University Olomouc. We thank J. Provazn\'{i}k for maintaining the cluster and providing support.

\section*{Disclosures}

The authors declare no conflict of interest.

\section*{Funding}

This work was supported by the Czech Science Foundation (project 21-18545S), and the Ministry of Education, Youth, and Sports of the Czech Republic (project OP JAC CZ.02.01.01/00/23\textunderscore021/0008790). DV acknowledges the support by Palack{\'y} University (projects IGA-PrF-2024-008, IGA-PrF-2025-010, and IGA-PrF-2026-005).\\


\bibliographystyle{naturemag}


\clearpage
\onecolumngrid

\section*{Supplementary Information for ``From Stars to Molecules: AI Guided Device-Agnostic Super-Resolution Imaging"}

\twocolumngrid

\subsection*{Experiment}

The optical setup, which provides resolution-limited experimental images with their corresponding ground truths, consists of three segments, see Extended Data Fig.~1. The first segment prepares the illumination of the digital micromirror device (DMD). An incoherent broadband light, produced by a laser-induced fluorescence light source (Crytur MonaLIGHT B01), is coupled to a 4~mm diameter light guide. The emerging light is collimated by a microscope objective $\text{MO}_{\text{s}}$ (OLYMPUS~20x/0.4). An iris diaphragm $\text{D}$ is positioned proximal to the objective. This diaphragm is re-imaged onto the DMD (Texas Instruments DLP LightCrafter 6500) by a telescope consisting of two achromatic doublets, $\text{L}_{\text{s1}}$ and $\text{L}_{\text{s2}}$, with focal lengths of 50~mm and 150~mm, respectively. The telescope improves the illumination homogeneity, and re-imaging the diaphragm reduces reflections from the passive parts of the DMD.

\begin{figure*}[]
	\renewcommand{\figurename}{Supplementary Fig.}
	\renewcommand{\thefigure}{1}
	\includegraphics[width=0.7\textwidth]{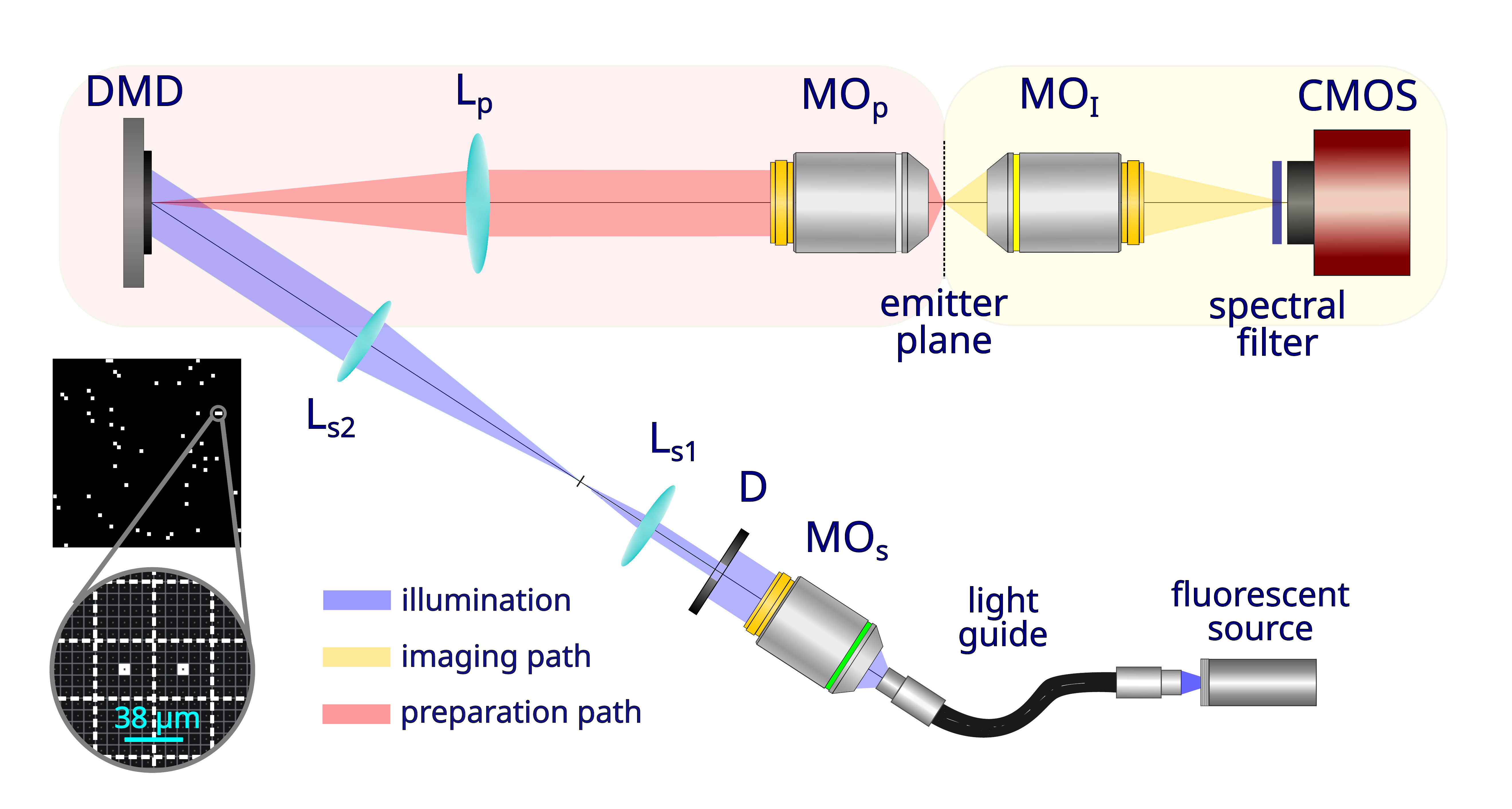}
	\caption{
		A detailed visualization of the optical setup with partial control and complete knowledge over the ground truth of emitters. In the illumination path (blue), we generate an incoherent light collimated by a microscopic objective $\text{MO}_{\text{s}}$. The subsequent iris diaphragm $\text{D}$ and two achromatic doublets, $\text{L}_{\text{s1}}$ and $\text{L}_{\text{s2}}$, improve the illumination homogeneity and reduce undesired reflections from the passive digital micromirror device DMD parts. In the preparation path (red), the DMD-induced ground-truth mask is re-imaged by an achromatic doublet $\text{L}_{\text{p}}$ and a high-resolution microscope objective $\text{MO}_{\text{p}}$, creating point-like emitters in the emitter plane. Lastly, this plane is projected by a low-resolution microscope objective $\text{MO}_{\text{I}}$ onto a CMOS camera with a mounted spectral filter in the imaging path (yellow).
	}
	\label{fig:7}
\end{figure*}

In the second setup segment, point-like emitters with the targeted spatial distribution are prepared. By addressing the DMD, we impose a ground-truth mask with each pixel comprising a $5 \times 5$ micromirror grid with a 7.6~$\mu$m micromirror pitch. As we address only the central micromirror of the grid, the closest distance between two emitting sources is 38~$\mu$m in the DMD plane, see Extended Data Fig.~1. This mask is re-imaged into the front focal plane of $\text{MO}_{\text{p}}$ by a preparation telescope consisting of a 200~mm focal-length achromatic doublet $\text{L}_{\text{p}}$ and a high-resolution microscope objective $\text{MO}_{\text{p}}$ (OLYMPUS~100x/0.9). As the point-like emitters, of the approximate 310~nm full width at half maximum and 340~nm minimal distance, are created in this plane, we term it the emitter plane.

Unlike the first two segments, which prepare the emitters and provide partial control and complete knowledge over their spatial distributions, the final segment represents an imaging setup with optical parameters unknown to the DAMN model. A low-resolution microscope objective $\text{MO}_{\text{I}}$ (OLYMPUS~10x/0.25) projects the emitter plane onto a CMOS camera (ZWO-ASI 178MM, 2.4~$\mu$m square pixels) with a mounted spectral filter (central wavelength of 532 nm with 10 nm spectral width). The $\text{MO}_{\text{I}}$ resolution is approximately 1.3~$\mu$m, regarding the Rayleigh limit in the emitter plane. In comparison, the 340~nm minimal emitter distance is almost 4~times below this resolution limit. The emitter plane is imaged with an approximate $230 \times$ magnification to match the major part of the camera chip. The exposure time for each measurement is set to 5.7~seconds to utilize the majority of the 14-bit dynamic range of the camera while avoiding its saturation. The captured image is down-scaled by a factor of 32 to match the $50 \times 50$ ground-truth mask size to enable metric evaluations. Using this imaging setup, we collected the experimental dataset of resolution-limited camera images and their corresponding ground-truth masks for various emitter concentrations, see Fig.~4~a of the main text. Additionally, we measured a high-SNR calibration set for precise estimation of the point spread function, which we provided only to both variants of the Richardson-Lucy algorithm and the Deep-STORM models. Besides the concentration ranging from 50 to 500 emitters, the optical parameters in the collected dataset exhibit the following estimated values: SNR~$= 2300$, average noise intensity~$= 10$, and an Airy-shaped kernel with $\sigma = 2.05$~px, equivalent to the 3.4~px full width at half maximum. The pixel-based Rayleigh resolution limit of our imaging system, given by the radius~R of the first dark ring in the Airy pattern, is approximately 3.9~px, see Fig.~4~b of the main text.


\subsection*{Aberration analysis}

\begin{figure*}[]
	\renewcommand{\figurename}{Supplementary Fig.}
	\renewcommand{\thefigure}{2}
	\includegraphics[width=0.8\textwidth]{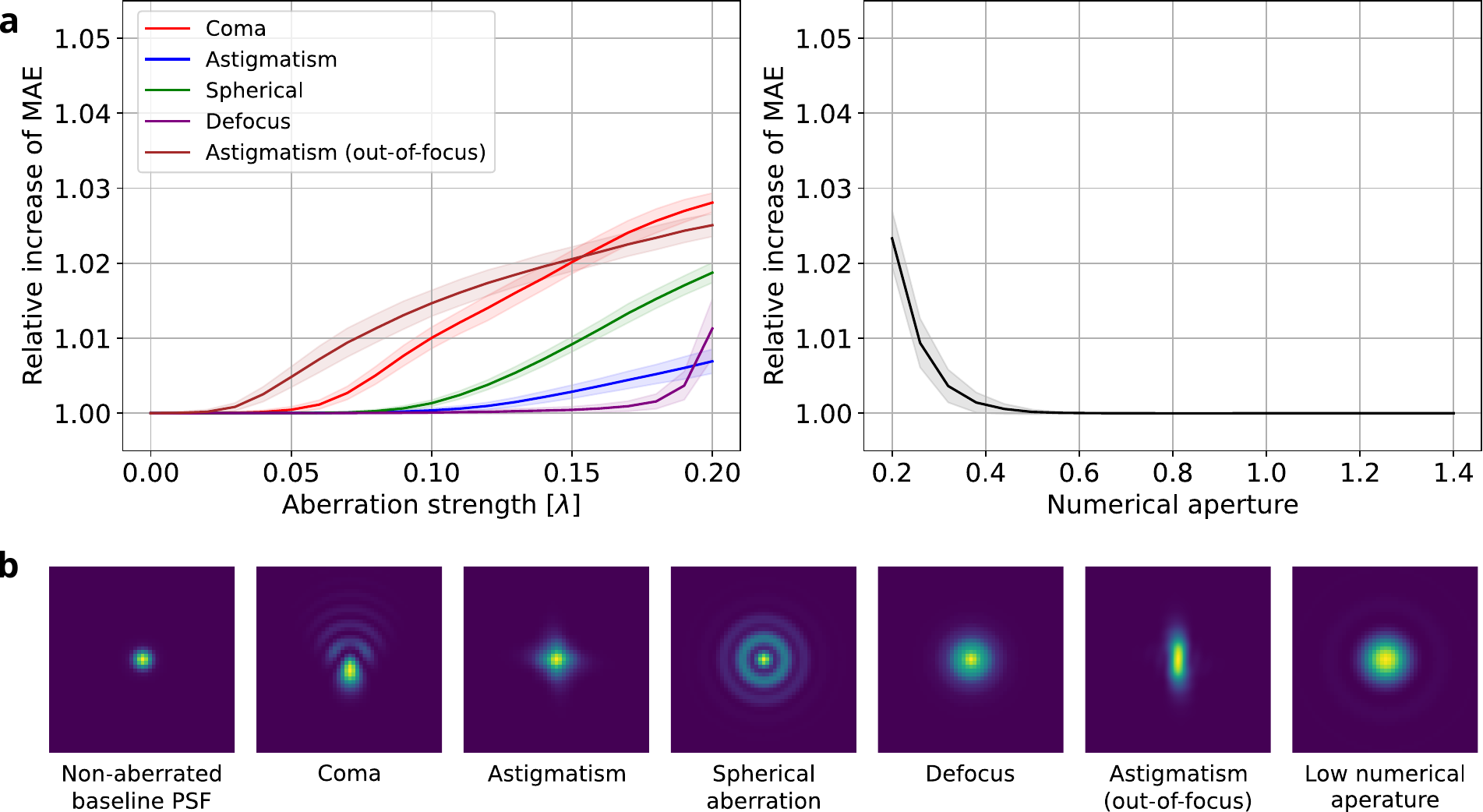}
	\caption{
		The aberration effects on the DAMN reconstruction performance. \textbf{(a)} The relative increase of mean absolute error (MAE) due to the numerical aperture (black) and aberration strength values for coma (red), astigmatism (blue), spherical aberration (green), defocus (purple), and out-of-focus astigmatism (brown). The presented results were evaluated relative to the baseline errors obtained for the non-aberrated point spread function (PSF). \textbf{(b)} Visualization of the distorted point spread functions for each aberration type with $0.2 \lambda$ strength, alongside the non-aberrated baseline and 0.2 numerical aperture.
	}
	\label{fig:8}
\end{figure*}

Since the DAMN model is trained on simulated data using either a Gaussian or an Airy point spread function, we evaluated its performance on images containing simulated optical aberrations. We generated vectorial point spread functions using a high-numerical-aperture Debye–Wolf diffraction model~\cite{Wolf1959}. Simulations assumed a wavelength of~$\lambda = 532$~nm, a numerical aperture of~1.4 (modified for the numerical-aperture dependence analysis), and a homogeneous immersion medium with a refractive index of~1.5. Circularly polarized illumination was assumed, and the vectorial pupil field included the standard $\sqrt{\text{cos}~\theta}$ apodization. Realistic aberrations of increasing strength were modeled using Zernike polynomials, specifically coma, astigmatism (both in- and out-of-focus), defocus, and spherical aberration. The aberration strength, referring to the value of the Zernike coefficient, is given in units of wavelength. For each aberration type and strength, we simulated 100 images with SNR~$= 10^3$, average noise intensity of~$10$, and~$20$ emitters. The results, including the dependence of reconstruction performance on numerical aperture, are shown in Extended Data Fig.~\ref{fig:8}. Any further details about the aberration simulation process can be found in our GitHub repository~\cite{Github}. The reported metric represents the relative change in mean absolute error compared to the baseline for non-aberrated point spread functions.

As illustrated in Extended Data Fig.~\ref{fig:8}~(a), the DAMN model demonstrates strong robustness against all aberration types. The relative error increase remains below 2\% across most realistic values of aberration strength and numerical aperture. Notably, gradually decreasing the numerical aperture shifts the intensity profile from a Gaussian to an Airy pattern while simultaneously broadening the effective width of the point spread function. Based on the results presented in Fig.~2 of the main text, the slight increase in relative error is primarily due to the increased width, which is approaching the field-of-view limits. Overall, these results indicate that the DAMN model reliably reconstructs even strongly aberrated images, with only a few percent increase in the mean absolute error metric.

Moreover, Extended Data Fig.~\ref{fig:8}~(b) shows the point spread functions for each aberration type at a strength of~$0.2 \lambda$, alongside the non-aberrated baseline and a point spread function with a numerical aperture of 0.2. These cases represent severely distorted point spread function profiles, which are rarely encountered in standard imaging scenarios. Nonetheless, the reconstruction errors are only marginally affected, and the trained model remains highly effective even for images with aberrations.

Finally, the reconstruction quality achieved in the biological (tubulin and nuclear pore complexes) and astronomical (double stars) applications further supports these findings. These experimentally acquired images inherently suffer from non-ideal imaging conditions and unknown aberrations. Nevertheless, the DAMN model reconstructs them with high fidelity, as discussed in the respective sections. Furthermore, the training data simulation could be extended to incorporate realistic aberrations, which would further enhance robustness to such imperfections at the cost of increased training process complexity. We deliberately refrained from doing so to demonstrate the strong generalization capability of the device-agnostic modeling, even when training on simplified simulation data.

\subsection*{Richardson-Lucy Algorithms}

The Richardson-Lucy algorithm is an iterative method used to restore an image blurred by a known point spread function~\cite{Richardson1972, Lucy1974}. This sophisticated and highly flexible algorithm is derived from the Bayesian probability theory with a flat prior. It maximizes the posterior probability that the observed image results from the estimated image convolved with the given point spread function. Assuming a multinomial distribution of detection events, the Richardson-Lucy is equivalent to the expectation-maximization algorithm for likelihood maximization in positive linear inverse problems~\cite{Dempster1977, VardiLee1993}. The quality of the reconstructed image relies heavily on accurate knowledge of the point spread function; discrepancies can lead to artifacts and inaccurate reconstructions. Since we simulate the testing data, the Richardson-Lucy algorithm can use the precise point spread function, making it an excellent method for providing a classical baseline for evaluating the DAMN model.

Moreover, the Richardson–Lucy algorithm can be viewed as an iterative maximum-likelihood method for Poisson observations. Under ideal model assumptions, if iterated to the maximum-likelihood solution and if the usual regularity conditions hold, its estimator may be asymptotically efficient, i.e., approach the classical Cramér–Rao bound in a large-data limit. This is the main reason we chose Richardson-Lucy as one of the benchmarking methods in our work.

In addition to the point spread function $P$, the iterative algorithm requires an initial guess for the reconstruction, $I^{(0)}$. A common practice is to set this guess as the observed resolution-limited image $I_{\text{blurred}}$ or to start with a uniform image. We explored both approaches and chose to set $I^{(0)} = I_{\text{blurred}}$, as there was a negligible difference in accuracy and computational speed. Then, an iterative procedure updates the estimation at the $k+1$ step
\begin{eqnarray}
	I^{\left( k+1 \right)} = I^{\left( k \right)} \cdot \left( \frac{I_{\text{blurred}}}{I^{\left( k \right)} \circledast P} \circledast P^* \right),
\end{eqnarray}
where $\circledast$ is convolution, and $P^*$ denotes the flipped point spread function. Both the initial guess $I^{(0)}$ and the point spread function $P$ need to be normalized to a unit sum. We implement two stopping criteria for the algorithm: either the mean absolute error between successive iterations becomes smaller than $10^{-10}$ per pixel, or the number of iterations exceeds $10^6$. The first criterion is typically met within $10^4$-$10^5$ iterations in the majority of cases.

Moreover, we evaluated a Richardson-Lucy algorithm with total variation regularization~\cite{Dey2006}. This variant is designed to suppress unstable oscillations and mitigate noise-amplification artifacts inherent to the classical formulation. In each iteration, the update step includes an additional division by the regularization term
\begin{eqnarray}
	1-\lambda_{TV} \nabla \cdot \left( \frac{\nabla I^{\left( k \right)}}{| \nabla I^{\left( k \right)} |} \right),
\end{eqnarray}
where $\lambda_{TV}=2 \times 10^{-3}$ denotes the regularization strength, $\nabla$~is the gradient operator, and $\nabla \cdot$~is the divergence operator. We tested several values of the regularization strength and chose to proceed with the original default value. The stopping criteria were kept identical to those of the non-regularized algorithm.
\\


\subsection*{Deep-STORM}

To benchmark our results also against a deep learning method, we implemented Deep-STORM~\cite{Nehme2018}, which produces super-resolution images from a dataset of stochastically blinking emitters. Deep-STORM is a convolutional neural network that reconstructs a high-resolution version of individual low-resolution frames. Therefore, it is directly comparable to our DAMN model. However, Deep-STORM is a device-specific method trained on simulated or experimentally acquired images with fixed imaging parameters, such as camera specification, point-spread-function profile, signal-to-noise ratio, and emitter density.

To compare its reconstructions in Fig.~2 and Fig.~4 of the main text, we trained numerous device-specific Deep-STORM models, one for each point in the graphs. These models were trained using simulated data with precise signal-to-noise ratios, concentrations, noise distributions, and point spread functions (widths and shapes), matching the targeted testing conditions. The model used to reconstruct the tubulin structures in Fig.~6 of the main text was trained by the authors and is available in their GitHub repository. This repository also includes the necessary code that we used to train all the remaining models. All training parameters, such as the loss function and the number of epochs, were kept intact as recommended by the authors.

Additionally, the Deep-STORM architecture inherently provides an eightfold increase in the reconstructed spatial dimensions. While we could deactivate this upsampling, it would significantly bottleneck the information flow throughout the architecture. Therefore, we chose to retain it to avoid disrupting the model capacity and complexity. Instead, we applied~$8 \times 8$ downscaling to the reconstructed images in places where the metric evaluation required non-upsampled image sizes (Fig.~2 and~4 of the main text).


\bibliographystyle{naturemag}


\end{document}